%% file: main.tex
\documentclass{article}

\usepackage{caption}
\usepackage{pdflscape} 
\usepackage{amsmath, amssymb, amsthm}
\usepackage{lmodern}
\usepackage{pgfplots}
\pgfplotsset{compat=1.18}
\usetikzlibrary{arrows.meta}
\usepackage[T1]{fontenc}
\usepackage{graphicx}
\usepackage{float}
\usepackage{tabularx}
\usepackage[margin=1in]{geometry}
\usepackage{booktabs}
\usepackage{natbib}
\usepackage{hyperref}
\usepackage[table]{xcolor}

\usetikzlibrary{positioning}

\title{Data \textit{(in)equities} in data science: Dissecting systemic and systematic biases in pulse oximetry}
\author{
  Lillian Rountree \qquad Harsh Parikh \qquad Bhramar Mukherjee \\[1.0em]
  Department of Biostatistics \\
  Yale University \\
  New Haven, CT, USA
}
\date{}

\begin{document}
\maketitle

\begin{abstract}
Data equity is an emerging framework for responsible data science. However, its core concepts, including fairness, representativeness, and information bias, remain largely abstract and general, lacking the mathematical specificity needed for practical implementation. In this paper, we demonstrate how statisticians can operationalize data equity by translating its tenets into precise, testable formulations tailored to a given problem. Using the well-documented case of differential measurement error across racial groups in pulse oximetry, we first adopt an oracle approach, tracing how a single upstream violation of information bias compounds through the analytic pipeline into treatment disparities, fairness violations, and adverse health outcomes. We then demonstrate the inverse: starting from an observed outcome disparity, the data equity framework provides a principled structure for systematically identifying its statistical sources. Our exposition underscores how data equity, prediction equity, and decision equity are distinct requirements with distinct evaluation and policy needs—a nuance that highlights both the unique role of statisticians in the era of artificial intelligence as well as the necessity of interdisciplinary collaboration.
\end{abstract}

\section{Introduction}

Algorithmic systems inform consequential decisions across social sciences, public health, law, business, and more. As valuable as these systems are, they can also result in unfavorable outcomes when the input data are treated as the unbiased truth. Myriad examples of how naive data use can perpetuate stereotypes and enable discrimination exist \citep{adam_mitigating_2022, zack_assessing_2024, bianchi_easily_2023}. 

These concerns have given rise to the concept of data equity, a wide-ranging term that encompasses all stages of an analytic workflow, from data collection and processing to final decision-making. To ground this broad idea, \citet{wang_ten_2026} outlined ten core concepts of data equity: fairness; accountability; transparency; ethics; privacy and confidentiality; selection bias; representativeness; generalizability; causality; and information bias. The framework provides a theoretical scaffold for data equity, but intentionally stops short of prescribing formal mathematical definitions. This was not an oversight but a deliberate reflection of the contextual nature of these tenets across a multitude of data settings and research questions. Careful mathematical thinking about study design, data collection, sources of biases, and the specific context of the problem at hand is needed to move from conceptual definitions to actionable equations. 

Such a principled approach to data is the hallmark of statistical training. Statisticians are uniquely positioned to tackle this crucial aspect of data equity in data science and artificial intelligence (AI), working in collaboration with domain experts and computer scientists to create appropriately designed, validated, and deployed inferential plans. This collaborative approach resists the impulse to immediately dive into model-building by taking data at face value and instead considers nuances of the data itself as much as the choice of model.

To demonstrate the usefulness of this statistical thinking in data science and AI, in this paper, we explore the well-known problem of differential measurement error in pulse oximetry. This stylized example demonstrates how information bias early in the analytic workflow leads to measurement error that accumulates downstream, resulting in unfair decisions and conclusions that are not generalizable. We use this example to illustrate how statistical clarity in these concepts enables identification of sources of data inequity, even when the root cause is unknown. We begin with an oracle perspective, demonstrating how an upstream error compounds into disparate health outcomes downstream. We then describe the inverse, demonstrating how statistics allows us to work from an observed data inequity to the identification of its source. In our conclusion, we point to the need for interdisciplinary collaboration to ensure data equity at all points in an analytic workflow and the necessity of foundational statistical thinking in the era of black box automation.

\subsection{Problem setting}

The differential measurement error of pulse oximetry is a textbook example of racial disparities in healthcare. The source of error is, on its own, an innocuous fact: the light absorption technology pulse oximeters use to measure oxygen saturation performs more poorly on tissue with high melanin levels (dark skin tones) than those with low levels. This disparity was well-characterized by the 1990s \citep{jubran_reliability_1990, cahan_agreement_1990}—over a decade after the pulse oximeter's rollout into clinical practice—but its potential clinical impact was ignored. After an influx of hospitalized patients on oxygen during the 2020 COVID-19 pandemic—which included a more enriched in-patient population of people of color—this disparity issue rose to the headlines of newspapers and scientific journals as the clinical consequences of this error became apparent \citep{rabin_pulse_2020}. In the high-profile paper by \citet{sjoding2020racial}, co-oximetry measurements in more than 10,000 patients across 178 medical centers revealed a systematic difference in the pulse oximeter-reported oxygen saturation levels compared with actual arterial oxygen saturation levels measured by a gold standard arterial blood gas (ABG) test. Alongside general measurement error for all patients, there was a strong difference in the errors across skin tones, with Black patients having, on average, 2\% lower arterial oxygen levels than white patients when their reported pulse oximeter readings were the same. \citet{sjoding2020racial} also showed that Black patients were nearly three times more likely than white patients (rates of 17\% vs 6\%) to experience ``occult'' or untreated hypoxemia, driven by erroneous pulse oximetry readings—meaning that this device error was indeed leading to actual undertreatment, downstream health consequences, and exacerbated existing racial disparities in healthcare \citep{gu_characteristics_2020}. 

As this problem is well-known and has a relatively well-understood mechanism of inequity, we choose to use it to demonstrate how data inequity refracts through different stages of a workflow. Specifically, we explore how measurements of oxygen saturation in an inpatient hospital setting impact the clinical decision to treat individuals with supplemental oxygen ($Z=1$) and downstream health outcomes $Y$. 

Pulse oximetry estimates arterial oxygen saturation by measuring differential light absorption through tissue. Let $W \in [0, 100]$ denote true arterial oxygen saturation, measured through arterial blood gas analysis, and let $W^*$ denote the noisier pulse oximeter estimate. Using a classic measurement error model \citep{carroll2006measurement}, we can portray the relationship as:

\begin{equation}\label{eq:measurement}
    W^* = W + \epsilon(A, X, W),
\end{equation}

where $\epsilon$ depends on skin pigmentation $A$ and other patient-level clinical factors $X$ (e.g., perfusion status). Crucially, $A$ has no actual impact on true arterial oxygen saturation $W$. As pulse oximetry is cheap, non-invasive, and fast, $W^*$ is often used as a proxy for $W$ in clinical settings. We consider a target population with distribution $P$ over $(X, A, Z, Y) \in \mathcal{X} \times \mathcal{A} \times \mathcal{Z} \times \mathcal{Y}$, where $X$ denotes covariates (including $W$ and other clinical factors $X_{-W}$), $A$ denotes a sensitive attribute, $Z$ denotes treatment, and $Y$ denotes the outcome. A selection mechanism $S \in \{0,1\}$ determines inclusion in the observed sample and $P(\cdot \mid S=1)$ denotes the study sample distribution. A decision rule $h:{\mathcal{W}}^* \rightarrow \mathcal{Z} = \{0,1\}$ maps the observed measurement $W^*$ to a treatment action (such as supplementary oxygenation through the nasal cannula). 

Throughout this paper, we consider a highly stylized version of the problem with pulse oximetry data. A rigorous exploration of the necessary assumptions and potential unmeasured confounders $U$ that may affect the measurement process and treatment decision is crucial for real-world analysis but omitted in the current presentation of the problem.

\section{Working forwards: Upstream measurement error leads to downstream data inequities}\label{oracle}

\subsection{Causality}

A thorough consideration of possible causal pathways—and thus possible confounders—is a crucial first step when approaching a data problem. A causal perspective provides us with language and methods to posit and then encode these assumed relationships between variables through tools such as directed acyclic graphs (DAGs). In the case of pulse oximetry, we assume that Figure \ref{dag} represents the truth. With this DAG, we can immediately identify the confounding pathway of skin pigmentation on the observed oxygen saturation ($W^*$) and its downstream reach.

\input{dag}

\subsection{Representativeness}

Representativeness is the most upstream data equity concern: without sufficient information across subgroups, downstream inequities cannot even be detected. In the case of pulse oximetry, \citet{jubran_reliability_1990} noted that calibration data were derived primarily from white individuals and recommended diversifying these samples to improve the device calibration across different skin pigmentation. Over thirty years later, \citet{tobin_pulse_2022} observed that their call to action had been largely ignored.

Representativeness is fundamentally about informational adequacy. Though a common necessary condition is distributional alignment between the study and target populations, $P(X, A \mid S=1) = P(X, A)$—often quantified as the participation-to-prevalence ratio $\text{PPR}_a := P(A=a \mid S=1)/P(A=a)$ \citep{eshera_demographics_2015}—this alone is not sufficient to guarantee that measures of model performance such as mean squared error (MSE) or confidence interval coverage will be equal across sensitive subgroups. To see why, consider estimation of the group-specific measurement bias, $\mu_a := \mathbb{E}[\epsilon \mid A=a]$. Detection of differential error requires estimating the contrast $\Delta := \mu_1 - \mu_0$, whose variance satisfies $ \text{Var}(\hat{\Delta}) \propto \frac{\sigma_1^2}{n_1} + \frac{\sigma_0^2}{n_0}$, where $n_a = n \cdot P(A=a \mid S=1)$ and $\sigma_a^2 = \text{Var}(\epsilon \mid A=a)$. When $P(A=1)$ is small, $n_1$ may be too small to detect $\Delta \neq 0$ even under proportional sampling. A stronger, information-theoretic sufficient condition requires that the Fisher information for each subgroup exceeds a detection threshold:

\begin{equation}\label{eq:fisher}
    I_{n_a}(\mu_a) \geq I^*, \quad \forall\, a \in \mathcal{A},
\end{equation}

where $I^*$ is obtained by inverting a power analysis at significance level $\alpha$ and power $1-\beta$:

\begin{equation}\label{eq:istar}
    I^* := \frac{(z_{\alpha/2} + z_\beta)^2}{\delta^2},
\end{equation}

and $\delta$ is the minimum clinically meaningful difference in $\mu_1 - \mu_0$. This links the representativeness criterion directly to the downstream fairness concern: $\delta$ encodes the smallest bias contrast that produces a consequential treatment disparity. Achieving $I_{n_1}(\mu_1) \geq I^*$ requires $n_1 \geq I^*\sigma_1^2$, which may exceed $nP(A=1 \mid S=1)$ under proportional sampling. In such cases, deliberate oversampling of underrepresented groups is necessary. This concern is closely related to questions in causal inference and generalizing causal effects; for more on these topics, consult \citet{parikh_2025, tipton}.

\subsection{Information bias}

Suppose that representativeness was adequately met for subgroups $A=a$ in the studies that evaluated pulse oximeters. For equitable measurement, we would require $\mathbb{E}[\epsilon \mid A] = c$ for all $A$—any systematic bias should be constant across groups. However, as several studies spanning decades have demonstrated, this condition does not hold for pulse oximetry \citep{jubran_reliability_1990, cahan_agreement_1990, bickler_effects_2005, sjoding2020racial}. Due to the light-based technology used to calculate oxygen saturation, the device inherently performs more poorly for darker skin tones. We can operationalize this phenomenon by letting $A=1$ denote dark skin tones compared with light ones ($A=0$)\footnote{It is important to recognize that casting skin pigmentation, a continuous variable, as binary is an oversimplification. While we could treat $A$ as continuous, we choose to proceed with $A$ as binary both for stylistic simplicity and to reflect the reality that, in practice, skin pigmentation is rarely measured directly. Instead of skin pigmentation, existing studies of differential measurement error in pulse oximetry often use race (typically only white and Black), despite this proxy being imperfect: two individuals of the same race may have very different levels of skin pigmentation, and a white/Black dichotomization excludes large swaths of the population. These are crucial considerations for real-world analysis of pulse oximetry data.}:

\begin{equation}\label{eq:infobias}
    \mathbb{E}[\epsilon \mid A = 1] > \mathbb{E}[\epsilon \mid A = 0] > 0.
\end{equation}

We note also that it has been observed in the experimental studies above that $\mathbb{E}[\epsilon(A,X,W) \mid A = 1]$ increases as the true oxygen level $W$ decreases—that is, the error of measurement is greatest where the clinical stakes are the highest. This phenomenon has also been observed clinically, as mentioned earlier: in one study, Black patients ($A=1$) experienced such ``occult hypoxemia'' ($W < 88\% $ but observed $W^* \geq 92 \%$) at nearly three times the rate of white patients ($A=0$) \citep{sjoding2020racial}. 

It is important to note that such measurement error, driven by observable variables, is not unresolvable. It is possible to measure and adjust for factors like skin tone that we know to influence measurement error, analogous to missing-at-random covariate adjustment approaches \citep{keogh2019_missingdata}. Of course, such adjustment approaches will only address the measurement error induced by covariates we observe, leaving the error driven by unseen factors like the underlying true oxygen saturation in place. Yet for pulse oximetry, even an adjustment for the observed source of measurement error was not implemented, leading to the downstream consequences we discuss below.

\subsection{Downstream disparities}

\paragraph{Treatment disparity.}

Clinical threshold rules based on reported oxygen levels $W^*$ determine treatment. Let $Z = \mathbf{1}(W^* < w_0)$ indicate clinical intervention, where $w_0$ is the protocol threshold (commonly $w_0 \in (88\%, 94\%)$). If the error distribution for $A=1$ is stochastically larger than that for $A=0$ at each $w < w_0$, consistent with the findings above, then what is clinically observed is:

\begin{equation}\label{eq:txdisparity}
    P(Z = 1 \mid W < w_0, A = 1) < P(Z = 1 \mid W < w_0, A = 0)
\end{equation}

\paragraph{Fairness.}

The treatment disparity of Equation \ref{eq:txdisparity} represents a fairness violation, specifically that of equality of opportunity\footnote{Operationalizing fairness has been a topic of research for many years, leading to a plethora of fairness metrics. Yet a crucial problem of fairness in algorithms is that many of these mathematical definitions directly contradict one another \citep{paulus_predictably_2020}. The decision of why fairness matters is problem-specific, underscoring the need to understand the context of the data and the desired use of the algorithm being developed.}. Equality of opportunity is met if, among individuals who truly require intervention, the probability of receiving it is independent of group membership:

\begin{equation}\label{eq:eqopp}
P(Z = 1 \mid W < w_0, A = a) = P(Z = 1 \mid W < w_0) \quad \text{for all } a \in \mathcal{A}.
\end{equation}

We can see that this is not met for pulse oximetry. With this differential measurement error, patients with darker skin ($A=1$) who truly have $W < w_0$ are less likely to have their condition detected and thus treated than those with lighter skin ($A=0$). 

\paragraph{Outcome disparity.}

Receiving supplemental oxygen ($Z=1$) can impact downstream health outcomes $Y$. If the differential measurement error and unequal odds for $Z=1$ are not identified and corrected, then under consistency ($Y = Y(Z)$) and conditional ignorability ($Y(z) \perp\!\!\!\perp Z \mid A$), the expected outcome for group $A=a$ decomposes to:

\begin{equation}\label{eq:potentialoutcome}
    \mathbb{E}[Y \mid A = a] = \mathbb{E}[Y(1) \mid A = a]P(Z = 1 \mid A = a) + \mathbb{E}[Y(0) \mid A = a]P(Z = 0 \mid A = a)
\end{equation}

The disparity attributable to differential measurement, assuming a constant beneficial treatment effect $\tau := \mathbb{E}[Y(0) \mid A=a] - \mathbb{E}[Y(1) \mid A=a] > 0$ (i.e., treatment reduces the adverse outcome) and no other biases, is:

\begin{equation}\label{eq:deltameas}
    \Delta_{\text{meas}} = \tau \cdot \left[ P(Z = 1 \mid A = 0) - P(Z = 1 \mid A = 1) \right]
\end{equation}

In the case of pulse oximetry, where $P(Z = 1 \mid A = 0) > P(Z = 1 \mid A = 1)$ due to differential measurement error, $\Delta_{\text{meas}} > 0$: differential measurement error produces worse health outcomes for group $A=1$.

\section{Working backwards: Identifying the source of data inequity}\label{backwards}

Starting from the initial sources of data inequity, we have demonstrated how, due to a lack of built-in checkpoints for inequity in the workflow, disparities compound throughout the analysis and use of pulse oximeters. Of course, this oracle approach is rarely feasible. Having outlined how data inequities filter through the system, we will now demonstrate how, by working backwards from a research question and referencing the ten core concepts of data equity, we can pinpoint sources of disparity. By considering each of the core concepts and its relevance to the problem at hand, we can begin to unpack sources of inequity.

Suppose we have health outcome data $Y$ from a hospital on patients admitted with acute respiratory illness. We observe differential health outcomes between Black and White patients and want to understand why this occurs at this specific hospital. A variety of health outcomes could be considered here, including days spent in the ICU, work days lost due to illness, or the need for mechanical ventilation (intubation). 

Assuming that being white or Black has no true biological influence on $Y$, there are two possible reasons for this discrepancy: 

\begin{enumerate}
    \item Systemic racial bias leads to differential treatment for patients based on $A=a$, despite otherwise identical measured health statuses.
    \item Systematic errors due to differences in the measurement of health status based on $A=a$ lead to differential treatment, despite identical true health statuses.
\end{enumerate}

We can formalize and test both of these conditions, which relate to the data equity concepts of \textit{fairness} and \textit{information bias}, respectively. If neither reason is supported by data analysis, then the assumption that there is not an underlying biological influence (for example, through genetic or physiological differences) may be investigated as a potential explanation.

\paragraph{Testing for systemic racial bias.}

Suppose we have sufficient representativeness to conduct statistically meaningful subgroup analyses on $A=a$ and measured conditions. We define systemic racial bias as the deviations from per-protocol treatment based on a variety of subjective (non-causal) information, including race, even if the measured condition is the same. Mathematically:

\begin{equation}\label{eq:systemic_null}
  Z \perp\!\!\!\perp A \mid W^*
\end{equation}

If systemic racial bias exists, then for some values of $W^*$:

\begin{equation}\label{eq:systemic_alt}
  P(Z = 1 \mid W^* = w^*,\, A = 1) \neq P(Z = 1 \mid W^* = w^*,\, A = 0)
\end{equation}

This is directly testable from the observed data $(W^*, A, Z)$, using non-parametric or parametric methods such as logistic regression or a Cochran-Mantel-Haenszel test.

\paragraph{Testing for differential measurement error.}

Suppose we find that systemic racial bias cannot fully explain the observed treatment disparity. We can then move to the second possibility of differential measurement error, as defined in Equation \ref{eq:infobias}. Testing this equation requires access to a gold standard measurement for a subset of patients, which may not always be feasible. However, the elimination of other sources of bias can provide evidence for the possibility that the information bias is due to measurement error and encourage the development of better measurement processes. In the case of pulse oximetry, for in-patient settings, the arterial saturation measurements are typically available, allowing us to compare the pulse oximeter readings to this gold standard. Given paired observations $(W^*, W, A)$ for a subset of patients, we can then test whether there is a difference in the distribution of ($W^*$, $W$) based on the value of $A$. One example of such an examination is provided by \citet{sjoding2020racial}, where the areas under the receiver operator characteristic curve (AUC) for correctly detecting $W<88\%$ based on $W^*$ were estimated for white and Black patients separately, revealing a statistically significant difference. 

\section{Illustration with synthetic pulse oximetry data}

Having outlined the propagation of bias both by working forwards and backwards, we now demonstrate how these biases can appear in practice. We construct a synthetic dataset with paired measurements of arterial oxygen saturation ($W$), pulse oximetry readings ($W^*$), race ($A$), treatment with supplemental oxygen ($Z$), and need for mechanical ventilation ($Y$). The data-generating process is based on the causal pathways of the DAG in Figure \ref{dag} and closely tracks the data depicted in \citet{sjoding2020racial}. It encodes two independent sources of bias: differential measurement error, where the pulse oximeter overreads more for Black patients ($A=1$) and increasingly does so at lower levels of true saturations, and systemic racial bias, where Black patients face lower odds of treatment even at the same measured value. 

\begin{figure}[H]
    \centering
    \includegraphics[width=0.7\linewidth]{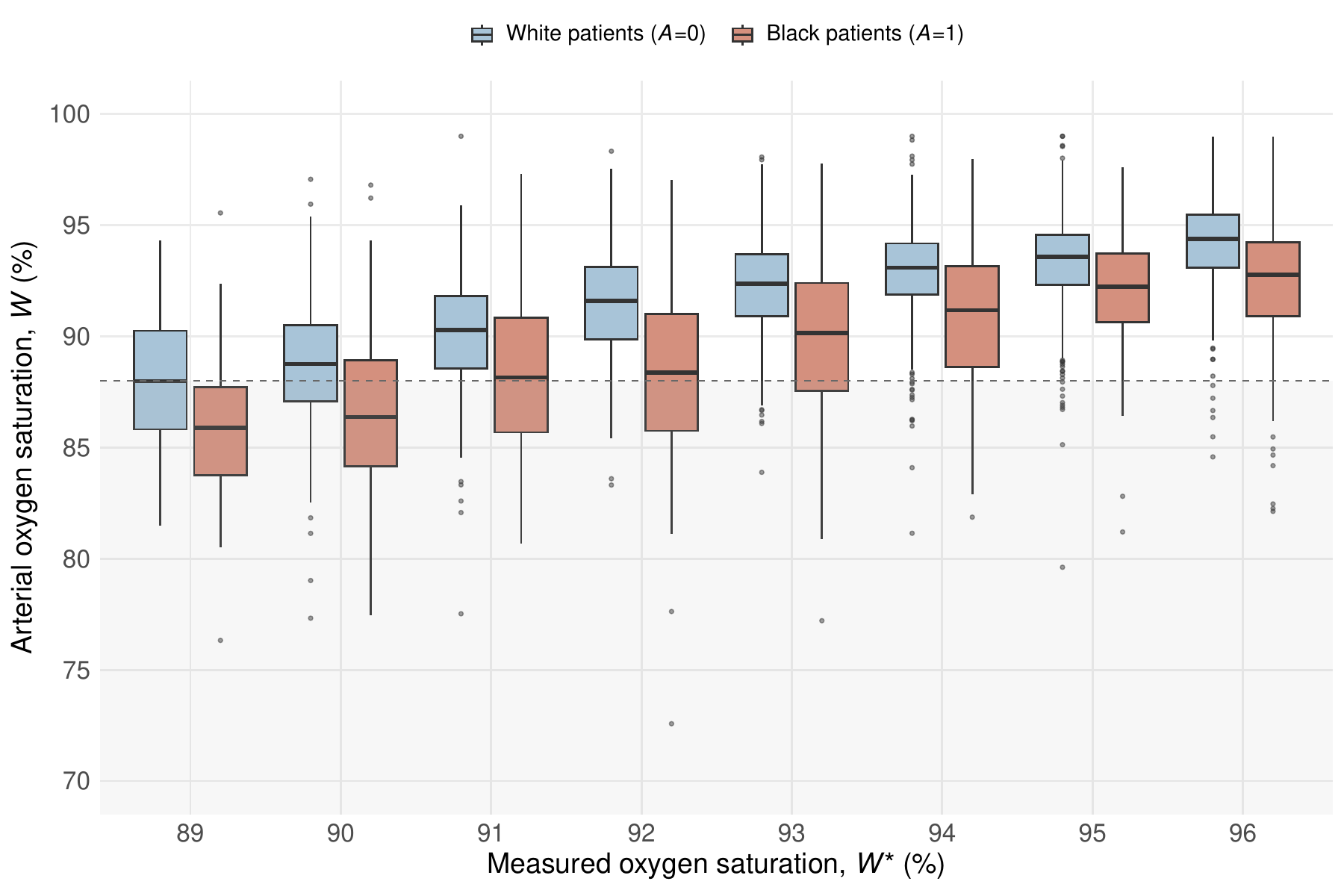}
    \caption{Pulse oximetry accuracy by race. Each box plot  shows the distribution of true arterial oxygen saturation ($W$) at each pulse oximetry reading ($W^*$), stratified by race. The dashed line marks the hypoxemia threshold ($W \leq 88\%$).}
    \label{fig:synthetic}
\end{figure}

Figure \ref{fig:synthetic} visualizes the differential measurement error of the pulse oximeter for white and Black patients at different oximeter readings. Table \ref{tab:hypoxemia_outcomes} shows how, when a deterministic treatment decision rule is used, this measurement error alone contributes to lower treatment rates for Black patients: 24\% of hypoxemic Black patients go untreated compared to 7\% of hypoxemic white patients in our synthetic dataset. When this measurement error compounds with systemic racial bias, we see a visible downstream impact on health outcomes. While systemic bias alone leads to a higher rate of mechanical ventilation in Black patients (8.4\%) compared to white patients (5.8\%) even when patients are treated based on their true oxygen saturation $W$, this gap in health outcomes is amplified by differential measurement error: when both systemic and systematic biases are at play, though the rate of ventilation for white patients remains fairly constant (5.9\%), the rate for Black patients jumps by nearly 5\%, up to 13.4\%. 

By using a synthetic dataset, we are able to further explore not just how differential measurement error alone impacts health, but also how it interacts with and compounds inequities stemming from systemic racial biases by toggling each source of bias on and off independently within the data-generating process. Table \ref{tab:metrics_table} demonstrates the consequences of these two sources of bias by calculating the metrics described Sections \ref{oracle} and \ref{backwards} for four datasets: both biases active, measurement error only, systemic racial bias only, and neither. All four datasets pass the representativeness threshold, confirming sufficient sample sizes to conduct the downstream analyses. By isolating each process, we demonstrate how measurement error alone produces significant differential error and marginal treatment disparities, while systemic racial bias alone produces significant treatment disparities and significant outcome disparities. Together, these negative effects are amplified: looking at our primary synthetic dataset used for Table \ref{tab:hypoxemia_outcomes} and Figure \ref{fig:synthetic}, which has both biases active, we see that almost all of the data equity metrics are flagged in Table \ref{tab:metrics_table}. In this case, there is statistically significant treatment disparity, equality of opportunity violations, and both parametric and non-parametric evidence of systemic racial bias, just as observed in real-world studies \citep{sjoding2020racial}.  When neither bias is active, no metric is flagged, confirming that these metrics do not generate spurious findings in the absence of true disparities.

\begin{table}[ht]
\centering
\begin{tabular}{lc|cc}
 & \textbf{Untreated hypoxemia} & \multicolumn{2}{c}{\textbf{Mechanical ventilation}} \\ \noalign{\smallskip}
 & $W < 88\%$ and $Z(W^*)=0$  & $Y(Z, W^*)=1$ & $Y(Z, W)=1$ \\
\noalign{\smallskip} \hline \noalign{\smallskip}
White patients ($A{=}0$) & 7.0\% & 5.9\% & 5.8\% \\ \noalign{\smallskip}
Black patients ($A{=}1$) & 24.0\% & 13.4\% & 8.4\% \\
\noalign{\smallskip}
\hline
\end{tabular}
\caption{Rates of undiagnosed hypoxemia and mechanical ventilation ($Y$) among white $(A = 0)$ and Black $(A = 1)$ patients under a deterministic treatment rule, where $Z$ indicates treatment with supplemental oxygen, $W$ indicates true oxygen saturation, and $W^*$ indicates measured saturation with a pulse oximeter. Comparing  columns $Y(Z, W^*) = 1$ and $Y(Z, W) = 1$ isolates the effect of differential measurement error; comparing rows within $Y(Z, W) = 1$ isolates the effect of systemic racial bias against Black patients.}
\label{tab:hypoxemia_outcomes}
\end{table}

\input{synthetic_data_table}

\section{Conclusion}

From the overarching idea of data equity, one can develop clear, context-specific approaches to mathematically assess its key tenets. Yet there is no one-size-fits-all prescriptive formulation for data equity; what is fair in one setting may not be in another, depending on the question of interest. For example, consistent differential measurement error across subgroups may still yield an adequate prediction in the overall population even as it produces disparate health outcomes when used to guide clinical decisions. In the case of pulse oximetry, however, \citet{sjoding2020racial} demonstrated significant differences in AUC for detecting hypoxemia with a pulse oximetry reading between white and Black patients, revealing predictive disparity alongside decision disparity. This demonstrates that data equity does not automatically equate to prediction equity, and prediction equity does not automatically equate to decision equity, three concepts we define and distinguish in Table \ref{tab:equity_def}. We also want to emphasize that, while our focus is on differential measurement error across known subgroups, all types of information bias run the risk of generating downstream inequities between subgroups, including missing data mechanisms \citep{wang_ten_2026, getzen_mining_2023}. Such nuances underscore the importance of statisticians in the development and implementation of algorithms and the interconnectedness of systematic and systemic biases.

What is also made clear in the pulse oximetry example is that it is not enough to identify the statistical problem in a data workflow. Differential measurement error by skin pigmentation for pulse oximetry was identified through lab studies not long after the device came onto the market, but a lack of the core data equity concepts of \textit{accountability} and \textit{transparency} stymied real change and awareness. It was only during the COVID-19 pandemic, where disproportionately high rates of hospitalization for Black Americans provided the clinical data needed to identify the problem in a real-world setting, that the consequences of this measurement error were broadly recognized \citep{rabin_pulse_2020}. Even now, tangible fixes remain limited: though the FDA recently released new regulatory guidance on pulse oximeters specifically to ensure ``non-disparate performance'' across skin pigmentation levels \citep{fda_pulse_2025}, this guidance is not legally binding, not without its critics \citep{lipnick_pulse_2025}, and, depending on how non-disparate performance is operationalized (by, for example, introducing different oxygen saturation thresholds for different races), it may itself open new pathways for disparity. 

No single discipline's focus alone is sufficient to prevent such outcomes. As the case of pulse oximetry makes clear, ensuring that algorithms reduce rather than perpetuate disparities will require sustained collaboration across statistics, computer science, clinical medicine, ethicists/philosophers, regulatory bodies, manufacturing corporations, the patients/consumers and health policy experts—and a system-level will to act on data-driven evidence.

\begin{table}[h]
\centering
\begin{tabular}{@{}lp{0.75\textwidth}@{}}
\toprule \noalign{\smallskip}
\textbf{Term} & \textbf{Definition} \\
\noalign{\smallskip}
\midrule
\noalign{\smallskip}
Data equity & The principle that data collection, analysis, and interpretation should be conducted in ways that minimize bias and maximize utility for everyone in the population. It is both a process and an outcome. \\[1ex]
\noalign{\smallskip}
Prediction equity & The principle that the identification and optimization of a predictive metric (such as AUC or Brier score) aligns with the predictive goal (eg. discrimination or accuracy) and performs equally for subgroups of interest in the training, testing, and target deployment setting.  \\[1ex]
\noalign{\smallskip}
Decision equity & The principle that everyone in the population has the agency to make informed decisions about their health (or other outcomes), based on recommendations supported by data and predictions that are representative and equitable.
\\
\noalign{\smallskip}
\bottomrule \\
\end{tabular} 
\caption{Our definitions of data equity, prediction equity, and decision equity. Based on \citet{wang_ten_2026}.}
\label{tab:equity_def}
\end{table}  

\bibliographystyle{apalike}
\bibliography{references}

\end{document}

%% file: dag.tex
\begin{figure}[htbp]
  \centering

\begin{tikzpicture}[
     bluenode/.style={rectangle, draw=black, fill=blue!10, minimum width=3cm, minimum height=1.2cm, align=center, font=\small},
    orangenode/.style={rectangle, draw=black, fill=orange!10, minimum width=3cm, minimum height=1.2cm, align=center, font=\small},
    arrow/.style={-{Stealth[length=3mm]}, thick}
]

\node[bluenode] ({True}) at (-6,0) {Arterial oxygen\\ saturation ($W$)};
\node[bluenode] ({Measured}) at (-2,0) {Measured oxygen\\ saturation ($W^*$)};
\node[bluenode] ({Outcome}) at (6, 0) {Treated with \\supplemental \\ oxygen ($Z$)};
\node[bluenode] ({Exposure}) at (2, 0) {Hypoxemia \\ when admitted \\ to hospital};
\node[bluenode] ({Survival}) at (6, -3) {Health outcome ($Y$)};

\node[orangenode] ({Skin}) at (-2, 3) {Skin\\ pigmentation $(A)$};
\node[orangenode] ({Bias}) at (3, 3) {Systemic biases};

\draw[arrow] ({True}) -- ({Measured});
\draw[arrow] ({Measured}) -- ({Exposure});
\draw[arrow] ({Exposure}) -- ({Outcome});
\draw[arrow] ({Skin}) -- ({Measured});
\draw[arrow] ({Skin}) -- ({Bias});
\draw[arrow] ({Bias}) -- ({Outcome});
\draw[arrow] (Bias.east) to[bend left=60] (Survival.east);
\draw[arrow] ({True}) to[bend right=20] ({Exposure});
\draw[arrow] ({True}.south) -- ({Survival});
\draw[arrow] ({Outcome}) -- ({Survival});
\draw[arrow] ({Exposure}) -- ({Survival});

\end{tikzpicture}
    \caption{DAG for the process determining a patient's outcome upon admission to the hospital, with the orange pathway indicating the confounders. Systemic biases encompass a variety of biases that could affect treatment decisions, regardless of the true health status, such as the subjective clinician impression of a patient, including  racial stereotypes. Unmeasured confounders $U$ (e.g., comorbidities) are omitted from the DAG for visual clarity but are assumed to impact all nodes. }
  \label{dag}
\end{figure}

%% file: synthetic_data_table.tex
\begin{landscape}
\begin{table}[htbp]
\centering
\scriptsize
\renewcommand{\arraystretch}{1.3}
\newcolumntype{P}[1]{>{\raggedright\arraybackslash}p{#1}}
\begin{tabular}{@{}P{4.25cm} P{4cm} P{4cm} P{2.5cm} P{2.5cm} P{2.5cm} P{2.5cm} @{}}
\toprule
\textbf{Metric} & \textbf{Description} & \textbf{Interpretation} & \textbf{Both} & \textbf{Measurement only} & \textbf{Systemic only} & \textbf{None} \\
\midrule
\multicolumn{7}{@{}l}{\textbf{Representativeness (Equations \ref{eq:fisher}--\ref{eq:istar})}} \\
\hline
$I_{n_a}(\mu_a) = \frac{n_a}{\sigma^2_a} \geq I^*$ & Fisher information for each subgroup (Detection threshold $I^* = 7.8$ when $\alpha = 0.05$, power$ = 0.80$, $\delta = 1\%$). & If $I_{n_a}(\mu_{a})>I^*$, there is sufficient data representativeness to identify disparities. & \parbox[t]{2.2cm}{$A{=}1$: 81.4 $>$ 7.8 \\ $A{=}0$: 653.3 $>$ 7.8} & \parbox[t]{2.2cm}{$A{=}1$: 81.4 $>$ 7.8 \\ $A{=}0$: 653.3 $>$ 7.8} & \parbox[t]{2.2cm}{$A{=}1$: 167.6 $>$ 7.8 \\ $A{=}0$: 653.3 $>$ 7.8} & \parbox[t]{2.2cm}{$A{=}1$: 167.6 $>$ 7.8 \\ $A{=}0$: 653.3 $>$ 7.8} \\
\midrule
\multicolumn{7}{@{}l}{\textbf{Information bias (Equation \ref{eq:infobias})}} \\
\hline
$\mathbb{E}[\epsilon \mid A{=}1] > \mathbb{E}[\epsilon \mid A{=}0] > 0$ & Mean measurement error by group, with the p-value from a one-sided t-test comparing group means. & If $p<0.01$, the measurement error is significantly larger for the $A{=}1$ subgroup. & \colorbox{yellow!30}{\parbox[t]{2.2cm}{$A{=}1$: 3.3\% \\ $A{=}0$: 1.3\% \\ $p{=}$$<$2$\times 10^{-16}$}} & \colorbox{yellow!30}{\parbox[t]{2.2cm}{$A{=}1$: 3.3\% \\ $A{=}0$: 1.3\% \\ $p{=}$$<$2$\times 10^{-16}$}} & \parbox[t]{2.2cm}{$A{=}1$: 1.4\% \\ $A{=}0$: 1.3\% \\ $p{=}$2.4$\times 10^{-2}$} & \parbox[t]{2.2cm}{$A{=}1$: 1.4\% \\ $A{=}0$: 1.3\% \\ $p{=}$2.4$\times 10^{-2}$} \\
\midrule
\multicolumn{7}{@{}l}{\textbf{Treatment disparity and fairness (Equations \ref{eq:txdisparity}--\ref{eq:eqopp})}} \\
\hline
$P(Z{=}1 \mid W{<}w_0, A{=}a)$ & Treatment rates among truly hypoxemic patients, with the p-value from a one-sided test of proportions for if $A{=}1$ patients are treated at a lower rate. & If $p<0.01$, truly hypoxemic $A{=}1$ individuals are treated at a significantly lower rate. & \colorbox{yellow!30}{\parbox[t]{2.2cm}{$A{=}1$: 64.9\% \\ $A{=}0$: 82.5\% \\ $p{=}$2.1$\times 10^{-5}$}} & \parbox[t]{2.2cm}{$A{=}1$: 73.9\% \\ $A{=}0$: 79.5\% \\ $p{=}$1.0$\times 10^{-1}$} & \colorbox{yellow!30}{\parbox[t]{2.2cm}{$A{=}1$: 68.6\% \\ $A{=}0$: 83.4\% \\ $p{=}$5.5$\times 10^{-3}$}} & \parbox[t]{2.2cm}{$A{=}1$: 82.9\% \\ $A{=}0$: 83.0\% \\ $p{=}$5.0$\times 10^{-1}$} \\
\hline
$P(Z{=}1 \mid W{<}w_0, A{=}a) - P(Z{=}1 \mid W{<}w_0)$ & Deviation from marginal treatment rate. $\chi^2$ test of independence and associated p-value. & If $p<0.01$, then equality of opportunity is violated: among the truly hypoxemic, treatment depends on the subgroup. & \colorbox{yellow!30}{\parbox[t]{2.2cm}{$A{=}1$: -9.2\% \\ $A{=}0$: 8.4\% \\ Marginal: 74.1\% \\ $\chi^2{=}$16.82 $p{=}$4.1$\times 10^{-5}$}} & \parbox[t]{2.2cm}{$A{=}1$: -2.9\% \\ $A{=}0$: 2.7\% \\ Marginal: 76.8\% \\ $\chi^2{=}$1.60 $p{=}$2.1$\times 10^{-1}$} & \parbox[t]{2.2cm}{$A{=}1$: -11.4\% \\ $A{=}0$: 3.5\% \\ Marginal: 79.9\% \\ $\chi^2{=}$6.46 $p{=}$1.1$\times 10^{-2}$} & \parbox[t]{2.2cm}{$A{=}1$: -0.1\% \\ $A{=}0$: 0.0\% \\ Marginal: 82.9\% \\ $\chi^2{=}$0.00 $p{=}$10.0$\times 10^{-1}$} \\
\midrule
\multicolumn{7}{@{}l}{\textbf{Outcome disparity (Equation \ref{eq:deltameas})}} \\
\hline
$P(Z{=}1 \mid A{=}0) - P(Z{=}1 \mid A{=}1)$ & Difference in probability of treatment between subgroups. $\chi^2$ test of independence for $Z \times A$. & If $p<0.01$, treatment rates differ significantly between subgroups. & \colorbox{yellow!30}{\parbox[t]{2.2cm}{0.1271 \\ $\chi^2{=}$44.41 \\ $p{=}$2.7$\times 10^{-11}$}} & \parbox[t]{2.2cm}{-0.0024 \\ $\chi^2{=}$0.01 \\ $p{=}$9.3$\times 10^{-1}$} & \colorbox{yellow!30}{\parbox[t]{2.2cm}{0.1448 \\ $\chi^2{=}$57.84 \\ $p{=}$2.8$\times 10^{-14}$}} & \parbox[t]{2.2cm}{0.0268 \\ $\chi^2{=}$1.90 \\ $p{=}$1.7$\times 10^{-1}$} \\
\hline
$\Delta = \tau \left[P(Z{=}1 \mid A{=}0) - P(Z{=}1 \mid A{=}1)\right]$ & Health outcome disparity attributable to the total treatment gap between subgroups due to measurement error, systemic racial bias, or both. & The excess ventilation risk borne by $A=1$ patients due to the combined treatment gap. Significant only if the $\chi^2$ test for the treatment gap above has $p<0.01$. & \colorbox{yellow!30}{\parbox[t]{2.2cm}{0.0035 ($\tau{=}$0.0276)}} & \parbox[t]{2.2cm}{-0.0001 ($\tau{=}$0.0285)} & \colorbox{yellow!30}{\parbox[t]{2.2cm}{0.0045 ($\tau{=}$0.0309)}} & \parbox[t]{2.2cm}{0.0008 ($\tau{=}$0.0295)} \\
\hline
$P(Y{=}1 \mid A{=}1) - P(Y{=}1 \mid A{=}0)$ & Observed ventilation disparity between subgroups. $\chi^2$ test of independence for $Y \times A$. & If $p<0.01$, ventilation rates differ significantly between subgroups. & \colorbox{yellow!30}{\parbox[t]{2.2cm}{0.0510 \\ $\chi^2{=}$35.01 \\ $p{=}$3.3$\times 10^{-9}$}} & \colorbox{yellow!30}{\parbox[t]{2.2cm}{0.0495 \\ $\chi^2{=}$32.22 \\ $p{=}$1.4$\times 10^{-8}$}} & \colorbox{yellow!30}{\parbox[t]{2.2cm}{0.0260 \\ $\chi^2{=}$9.79 \\ $p{=}$1.8$\times 10^{-3}$}} & \parbox[t]{2.2cm}{0.0177 \\ $\chi^2{=}$4.42 \\ $p{=}$3.5$\times 10^{-2}$} \\
\midrule
\multicolumn{7}{@{}l}{\textbf{Systemic racial bias (Equations \ref{eq:systemic_null}--\ref{eq:systemic_alt})}} \\
\hline
$Z \perp\!\!\!\perp A \mid W^*$ & Logistic regression of treatment on measured oxygen saturation and subgroup membership: $Z \sim W^* + A$. & When $p<0.01$, treatment decisions depend on subgroup membership even after conditioning on $W^*$, suggesting bias beyond measurement error. & \colorbox{yellow!30}{\parbox[t]{2.2cm}{$\hat{\beta}_{A}{=}$-0.655 \\ $p{=}$1.0$\times 10^{-13}$}} & \parbox[t]{2.2cm}{$\hat{\beta}_{A}{=}$0.014 \\ $p{=}$8.7$\times 10^{-1}$} & \colorbox{yellow!30}{\parbox[t]{2.2cm}{$\hat{\beta}_{A}{=}$-0.756 \\ $p{=}$$<$2$\times 10^{-16}$}} & \parbox[t]{2.2cm}{$\hat{\beta}_{A}{=}$-0.142 \\ $p{=}$1.1$\times 10^{-1}$} \\
\hline
$Z \perp\!\!\!\perp A \mid W^*$ & Cochran-Mantel-Haenszel test for conditional independence of treatment and subgroup, stratified by $W^*$. & If $p<0.01$, the association between treatment and subgroup persists across $W^*$ levels, providing non-parametric evidence of bias beyond measurement error. & \colorbox{yellow!30}{\parbox[t]{2.2cm}{$p{=}$7.6$\times 10^{-14}$}} & \parbox[t]{2.2cm}{$p{=}$9.1$\times 10^{-1}$} & \colorbox{yellow!30}{\parbox[t]{2.2cm}{$p{=}$$<$2$\times 10^{-16}$}} & \parbox[t]{2.2cm}{$p{=}$1.1$\times 10^{-1}$} \\
\bottomrule
\end{tabular}
\caption{Data equity metrics calculated on synthetic datasets generated under four bias scenarios: both measurement error and systemic racial bias, measurement error only, systemic racial bias only, and neither. Black patients are in subgroup $A=1$; white patients are in subgroup $A=0$. The health outcome $Y$ is the need for mechanical ventilation; hypoxemia is defined as $W < 88\%$; and the treatment $Z$ is supplemental oxygen. When appropriate, cells are \colorbox{yellow!30}{highlighted} to indicate a significant result ($p<0.01$). Details of the data-generating process are available on \href{https://github.com/lirountree/dataequity_cjs/tree/main}{GitHub}.}
\label{tab:metrics_table}
\end{table}
\end{landscape}

%% file: references.bib
@article{wang_ten_2026,
	title = {Ten {Core} {Concepts} for {Ensuring} {Data} {Equity} in {Public} {Health}},
	volume = {7},
	issn = {2689-0186},
	url = {https://doi.org/10.1001/jamahealthforum.2025.6031},
	doi = {10.1001/jamahealthforum.2025.6031},
	number = {1},
	urldate = {2026-01-09},
	journal = {JAMA Health Forum},
	author = {Wang, Yiran and Boyd, Alicia E. and Rountree, Lillian and Ren, Yi and Nyhan, Kate and Nagar, Ruchit and Higginbottom, Jackson and Ranney, Megan L. and Parikh, Harsh and Mukherjee, Bhramar},
	month = jan,
	year = {2026},
	pages = {e256031--e256031},
}

@article{sjoding2020racial,
  title={Racial bias in pulse oximetry measurement},
  author={Sjoding, Michael W and Dickson, Robert P and Iwashyna, Theodore J and Gay, Steven E and Valley, Thomas S},
  journal={New England Journal of Medicine},
  volume={383},
  number={25},
  pages={2477--2478},
  year={2020},
  publisher={Massachusetts Medical Society},
  url = {https://www.nejm.org/doi/10.1056/NEJMc2029240}
}

@article{bickler_effects_2005,
	title = {Effects of {Skin} {Pigmentation} on {Pulse} {Oximeter} {Accuracy} at {Low} {Saturation}},
	volume = {102},
	issn = {1528-1175},
	url = {https://journals.lww.com/anesthesiology/fulltext/2005/04000/effects_of_skin_pigmentation_on_pulse_oximeter.3.aspx},
	number = {4},
	journal = {Anesthesiology},
	author = {Bickler, Philip E. and Feiner, John R. and Severinghaus, John W.},
	year = {2005},
}

@article{jubran_reliability_1990,
	title = {Reliability of {Pulse} {Oximetry} in {Titrating} {Supplemental} {Oxygen} {Therapy} in {Ventilator}-{Dependent} {Patients}},
	volume = {97},
	issn = {0012-3692},
	url = {https://www.sciencedirect.com/science/article/pii/S0012369216320293},
	doi = {10.1378/chest.97.6.1420},
	number = {6},
	journal = {Chest},
	author = {Jubran, Amal and Tobin, Martin J.},
	month = jun,
	year = {1990},
	pages = {1420--1425},
}

@article{paulus_predictably_2020,
	title = {Predictably unequal: understanding and addressing concerns that algorithmic clinical prediction may increase health disparities},
	volume = {3},
	issn = {2398-6352},
	url = {https://doi.org/10.1038/s41746-020-0304-9},
	doi = {10.1038/s41746-020-0304-9},
	number = {1},
	journal = {npj Digital Medicine},
	author = {Paulus, Jessica K. and Kent, David M.},
	month = jul,
	year = {2020},
	pages = {99},
}

@article{eshera_demographics_2015,
	title = {Demographics of {Clinical} {Trials} {Participants} in {Pivotal} {Clinical} {Trials} for {New} {Molecular} {Entity} {Drugs} and {Biologics} {Approved} by {FDA} {From} 2010 to 2012},
	volume = {22},
	issn = {1075-2765},
	url = {https://journals.lww.com/americantherapeutics/fulltext/2015/11000/demographics_of_clinical_trials_participants_in.6.aspx},
	number = {6},
	journal = {American Journal of Therapeutics},
	author = {Eshera, Noha and Itana, Hawi and Zhang, Lei and Soon, Greg and Fadiran, Emmanuel O.},
	year = {2015},
}

@article{tobin_pulse_2022,
	title = {Pulse oximetry, racial bias and statistical bias},
	volume = {12},
	issn = {2110-5820},
	url = {https://doi.org/10.1186/s13613-021-00974-7},
	doi = {10.1186/s13613-021-00974-7},
	number = {1},
	journal = {Annals of Intensive Care},
	author = {Tobin, Martin J. and Jubran, Amal},
	month = jan,
	year = {2022},
	pages = {2},
}

@article{rabin_pulse_2020,
	address = {New York},
	chapter = {A},
	title = {Pulse {Oximeter} {Devices} {Have} {Higher} {Error} {Rate} in {Black} {Patients}},
	url = {https://www.nytimes.com/2020/12/22/health/oximeters-covid-black-patients.html},
	journal = {New York Times},
	author = {Rabin, Roni Caryn},
	month = dec,
	year = {2020},
	pages = {4},
}

@book{carroll2006measurement,
  title={Measurement Error in Nonlinear Models: A Modern Perspective},
  author={Carroll, Raymond J and Ruppert, David and Stefanski, Leonard A and Crainiceanu, Ciprian M},
  year={2006},
  edition={2nd},
  publisher={Chapman \& Hall/CRC},
  address={Boca Raton, FL}
}

@article{adam_mitigating_2022,
    title = {Mitigating the impact of biased artificial intelligence in emergency decision-making},
    volume = {2},
    issn = {2730-664X},
    url = {https://doi.org/10.1038/s43856-022-00214-4},
    doi = {10.1038/s43856-022-00214-4},
    number = {1},
    journal = {Communications Medicine},
    author = {Adam, Hammaad and Balagopalan, Aparna and Alsentzer, Emily and Christia, Fotini and Ghassemi, Marzyeh},
    month = nov,
    year = {2022},
    pages = {149},
}

@article{zack_assessing_2024,
    title = {Assessing the potential of {GPT}-4 to perpetuate racial and gender biases in health care: a model evaluation study},
    volume = {6},
    issn = {2589-7500},
    url = {https://doi.org/10.1016/S2589-7500(23)00225-X},
    doi = {10.1016/S2589-7500(23)00225-X},
    number = {1},
    urldate = {2026-02-12},
    journal = {The Lancet Digital Health},
    publisher = {Elsevier},
    author = {Zack, Travis and Lehman, Eric and Suzgun, Mirac and Rodriguez, Jorge A and Celi, Leo Anthony and Gichoya, Judy and Jurafsky, Dan and Szolovits, Peter and Bates, David W and Abdulnour, Raja-Elie E and Butte, Atul J and Alsentzer, Emily},
    month = jan,
    year = {2024},
    pages = {e12--e22},
}

@inproceedings{bianchi_easily_2023,
    address = {New York, NY, USA},
    series = {{FAccT} '23},
    title = {Easily {Accessible} {Text}-to-{Image} {Generation} {Amplifies} {Demographic} {Stereotypes} at {Large} {Scale}},
    isbn = {979-8-4007-0192-4},
    url = {https://doi.org/10.1145/3593013.3594095},
    doi = {10.1145/3593013.3594095},
    booktitle = {Proceedings of the 2023 {ACM} {Conference} on {Fairness}, {Accountability}, and {Transparency}},
    publisher = {Association for Computing Machinery},
    author = {Bianchi, Federico and Kalluri, Pratyusha and Durmus, Esin and Ladhak, Faisal and Cheng, Myra and Nozza, Debora and Hashimoto, Tatsunori and Jurafsky, Dan and Zou, James and Caliskan, Aylin},
    year = {2023},
    pages = {1493--1504},
}

@article{tipton,
 ISSN = {10769986, 19351054},
 URL = {http://www.jstor.org/stable/43966357},
 author = {Elizabeth Tipton},
 journal = {Journal of Educational and Behavioral Statistics},
 number = {6},
 pages = {478--501},
 publisher = {[American Educational Research Association, Sage Publications, Inc., American Statistical Association]},
 title = {How Generalizable Is Your Experiment? An Index for Comparing Experimental Samples and Populations},
 urldate = {2026-06-15},
 volume = {39},
 year = {2014}
}

@misc{keogh2019_missingdata,
      title={Measurement error as a missing data problem}, 
      author={Ruth H. Keogh and Jonathan W. Bartlett},
      year={2019},
      eprint={1910.06443},
      archivePrefix={arXiv},
      primaryClass={stat.ME},
      url={https://arxiv.org/abs/1910.06443}, 
}

@article{parikh_2025,
   title={Who Are We Missing?: A Principled Approach to Characterizing the Underrepresented Population},
   volume={120},
   ISSN={1537-274X},
   url={http://dx.doi.org/10.1080/01621459.2025.2495319},
   DOI={10.1080/01621459.2025.2495319},
   number={551},
   journal={Journal of the American Statistical Association},
   publisher={Informa UK Limited},
   author={Parikh, Harsh and Ross, Rachael K. and Stuart, Elizabeth and Rudolph, Kara E.},
   year={2025},
   month=June, pages={1414–1423} }

@article{lipnick_pulse_2025,
    title = {Pulse {Oximetry} and {Skin} {Pigmentation}—{New} {Guidance} {From} the {FDA}},
    volume = {333},
    issn = {0098-7484},
    url = {https://doi.org/10.1001/jama.2025.1959},
    doi = {10.1001/jama.2025.1959},
    number = {16},
    urldate = {2026-02-18},
    journal = {JAMA},
    author = {Lipnick, Michael S. and Ehie, Odinakachukwu and Igaga, Elizabeth N. and Bicker, Philip},
    month = apr,
    year = {2025},
    pages = {1393--1395},
}

@techreport{fda_pulse_2025,
    address = {Washington, DC},
    title = {Pulse {Oximeters} for {Medical} {Purposes} - {Non}-{Clinical} and {Clinical} {Performance} {Testing}, {Labeling}, and {Premarket} {Submission} {Recommendations}},
    url = {https://www.fda.gov/regulatory-information/search-fda-guidance-documents/pulse-oximeters-medical-purposes-non-clinical-and-clinical-performance-testing-labeling-and},
    number = {FDA-2023-N-4976},
    institution = {Food and Drug Administration},
    author = {FDA},
    month = jan,
    year = {2025},
}

@article{getzen_mining_2023,
    title = {Mining for equitable health: {Assessing} the impact of missing data in electronic health records},
    volume = {139},
    issn = {15320464},
    shorttitle = {Mining for equitable health},
    url = {https://linkinghub.elsevier.com/retrieve/pii/S153204642200274X},
    doi = {10.1016/j.jbi.2022.104269},
    language = {en},
    urldate = {2026-06-25},
    journal = {Journal of Biomedical Informatics},
    author = {Getzen, Emily and Ungar, Lyle and Mowery, Danielle and Jiang, Xiaoqian and Long, Qi},
    month = mar,
    year = {2023},
    pages = {104269},
}

@article{cahan_agreement_1990,
    title = {Agreement between {Noninvasive} {Oximetric} {Values} for {Oxygen} {Saturation}},
    volume = {97},
    copyright = {https://www.elsevier.com/tdm/userlicense/1.0/},
    issn = {00123692},
    url = {https://linkinghub.elsevier.com/retrieve/pii/S0012369216350462},
    doi = {10.1378/chest.97.4.814},
    language = {en},
    number = {4},
    urldate = {2026-06-29},
    journal = {Chest},
    author = {Cahan, Clement and Decker, Michael J. and Hoekje, Peter L. and Strohl, Kingman P},
    month = apr,
    year = {1990},
    pages = {814--819},
}

@article{gu_characteristics_2020,
    title = {Characteristics {Associated} {With} {Racial}/{Ethnic} {Disparities} in {COVID}-19 {Outcomes} in an {Academic} {Health} {Care} {System}},
    volume = {3},
    issn = {2574-3805},
    url = {https://doi.org/10.1001/jamanetworkopen.2020.25197},
    doi = {10.1001/jamanetworkopen.2020.25197},
    number = {10},
    urldate = {2023-09-26},
    journal = {JAMA Network Open},
    author = {Gu, Tian and Mack, Jasmine A. and Salvatore, Maxwell and Prabhu Sankar, Swaraaj and Valley, Thomas S. and Singh, Karandeep and Nallamothu, Brahmajee K. and Kheterpal, Sachin and Lisabeth, Lynda and Fritsche, Lars G. and Mukherjee, Bhramar},
    month = oct,
    year = {2020},
    pages = {e2025197--e2025197},
}
